\documentclass[useAMS,usenatbib]{mn2e}              

\usepackage[dvips]{graphicx}
\title[Large-Scale Structure of LAEs]
{Large-Scale Structure of Short-Lived Lyman$\alpha$ Emitters}
\author[I Shimizu., M Umemura., and A Yonehara]{Ikkoh Shimizu,$^{1}$\thanks {E-mail:shimizu@ccs.tsukuba.ac.jp}
Masayuki Umemura,$^{1}$ \thanks {E-mail:umemura@ccs.tsukuba.ac.jp}
Atsunori Yonehara,$^{2}$ \thanks {E-mail: yonehara@cc.kyoto-su.ac.jp}
\thanks {JSPS Postdoctal Fellowships for Research Abroad}\\
$^{1}$Center for Computational Sciences, University of Tsukuba, Tsukuba 305-8577, Japan \\
$^{2}$ Astronomisches Rechen-Institut,
Zentrum f\"{u}r Astronomie der Universit\"{a}t Heidelberg,
M\"{o}nchhofstrasse 12-14, 69120, Heidelberg, Germany;\\
Department of Physics, Faculty of Science, Kyoto Sangyo University, 
    Motoyama, Kamigamo, Kita-ku, Kyoto, 603-8555, Japan}
\begin{document}

%\date{Accepted 2007 Month Date. Received 200X Month Date; in original form 2005 Month date}

\pagerange{\pageref{firstpage}--\pageref{lastpage}} \pubyear{2002}

\maketitle

\label{firstpage}

\begin{abstract}
Recently discovered large-scale structure of Ly$\alpha$ Emitters (LAEs)
raises a novel challenge to the cold dark matter (CDM) cosmology. 
The structure is extended over more than 50~Mpc at redshift $z=3.1$, 
and exhibits a considerably weak angular correlation. 
Such properties of LAE distributions appear to be incompatible with
the standard biased galaxy formation scenario in the CDM cosmology. 
In this paper, by considering the possibility that LAEs
are short-lived events, we attempt to build up the picture of LAEs
concordant with the CDM cosmology. 
We find that if the lifetime of LAEs is as short as
$(6.7 \pm 0.6) \times 10^7$~yr,
the distributions of simulated galaxies successfully match
the extension and morphology of large-scale structure of LAEs at $z=3.1$,
and also the weak angular correlation function.
This result implies that LAEs at $z=3.1$ do not necessarily reside in
high density peaks, but tends to be located in less dense regions,
in a different way from the expectation by the standard biased 
galaxy formation scenario. 
In addition, we make a prediction for the angular correlation function of LAEs 
at redshifts higher than 3. 
It is found that the prediction deviates from that by the standard biased 
galaxy formation scenario even at redshifts $4 \la z \la 6$. 
\end{abstract}


\section{Introduction}
Recently, the deep imaging surveys by $8 \sim 10 {\rm ~m}$ class telescopes with narrow-band filter have effectively revealed the properties of 
Ly$\alpha$ emitters (LAEs), which are one class of high redshift objects
\citep{Co98, Hu98, Hu99, Hu02}.
Based on the observational results, 
it is inferred that LAEs have smaller sizes, much less dust, and 
a smaller amount of stellar component 
than the other class of high redshift galaxies, e.g., Lyman Break Galaxies (LBGs) at same redshifts \citep{Sha01,Ven05}. 
The spatial distribution of observed LAEs generally shows large filamentary structure \citep{Shimasaku03, Ou04,Ou05, Matsuda05}. 
However, as pointed out by \citet{Ha04} recently, 
the observed properties of LAEs 
such as weak angular correlation function (ACF)
are not explained well by a standard biased galaxy formation scenario
in the context of $\Lambda$ cold dark matter ($\Lambda$CDM) cosmology.
Especially, the large-scale structure of LAEs found by
\citet{H04} is difficult to reproduce. 
The large-scale structure shows belt-like structure rather than filamentary structure, 
and may correspond to $6\sigma$ density fluctuation if it follows underlying dark matter distribution \citep{Kau99}.
ACF is significantly weaker than that predicted in
a conventional biased galaxy formation model (e.g. \citet{Kau97}).
Moreover, ACF becomes negative at small scale 
of $< 180$~arcsec ($<6h^{-1}$~Mpc) in high density regions (HDR) 
\citep{H04}. 
Hence, the observational features of spatial distribution of LAEs 
appear to be incompatible with the standard biased 
galaxy formation model.

From a theoretical point of view, it is recently argued 
that LAEs corresponds to an early chemodynamical evolution phase of
primordial galaxies \citep{MU06a, MU06b}. 
In an ultra high resolution simulation on the dynamical and chemical evolution 
of galaxy by \citet{MU06a, MU06b},
it is shown that multiple supernova explosions at an early phase of
$<3 \times 10^8$~yr result in forming high density cooling shells, 
which emit so strong Ly$\alpha$ as to account for the luminosity of LAEs.
%Since Ly$\alpha$ emission is easily absorbed by dust, 
%LAEs cannot be observed as LAEs after the metal enrichment has 
%proceeded in some degree. 
%Here, this period is regarded as the end of lifetime of LAEs.
However, it has not been argued whether this picture of LAEs 
is consistent with the observation. 

In this letter, the spatial distributions of LAEs are simulated
by taking into account the lifetime of the emitters,
which has not been thitherto considered in
the standard biased galaxy formation scenario \citep{Kau99, Ha04}.
Then, we investigate whether the picture of short-lived LAEs can
explain the clustering properties of LAEs found by \citet{H04}.
In \S2, we describe the basic picture and numerical method.
In \S3, the results are presented with some discussion. 
\S4 is devoted to the summary.
Throughout this letter, 
we adopt $\Lambda$CDM cosmology with the matter density 
$\Omega_{\rm{M}} = 0.3$, 
the cosmological constant $\Omega_{\Lambda} = 0.7$, 
the Hubble constant $h = 0.7$ in units of 
$H_0 = 100 \rm{~km ~s^{-1} ~Mpc^{-1}}$, 
the baryon density $\Omega_{\rm B}h^2 = 0.02$, and $\sigma_8 = 0.92$ \citep{WMAP}.

\section{Model}
\begin{figure}
\includegraphics[width = 85mm]{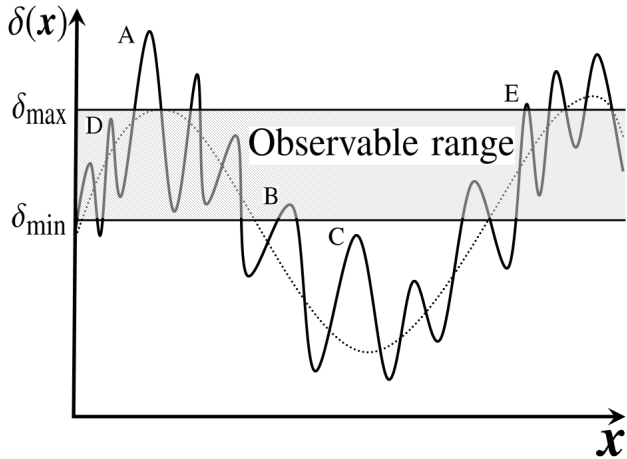}
\caption{Schematic view of the present model. 
The amplitude of density fluctuations at a redshift are
shown against the spatial location. 
The density peaks above $\delta_{\rm{max}}$, e.g., peak A, 
have already finished their lifetime as LAEs.
In contrast, the density peaks below $\delta_{\rm{min}}$ e.g., peak C, 
do not yet start to shine as LAEs.
The density peaks between $\delta_{\rm{min}}$ and $\delta_{\rm{max}}$, 
e.g., peaks B or D, can be observed as LAEs.
In this view, peak A is the oldest galaxy after LAE phase.
Peak B is the youngest LAEs and peak D is the oldest LAEs.
Peak E has just finished to shine as LAEs at the redshift.
}
\label{gainen}
\end{figure}

\begin{figure*}
\includegraphics[width = 170mm]{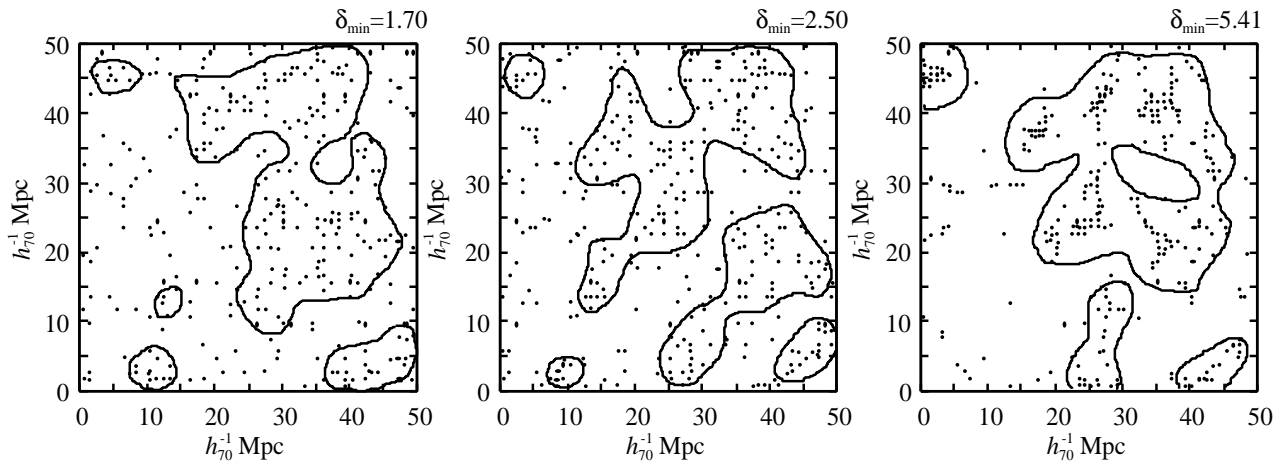}
\caption{LAE distribution for different values of $\delta_{\rm{min}}$
and $\delta_{\rm{max}}$.
Left panel is the LAE distribution for the model
with $\delta_{\rm{min}} = 1.7$ and $\delta_{\rm{max}} = 1.75$, and
middle panel is $\delta_{\rm{min}} = 2.5$ and $\delta_{\rm{max}} = 2.63$. 
Right panel is $\delta_{\rm{min}} = 5.4$ and $\delta_{\rm{max}} = \infty$,
which is corresponding to a conventional biased galaxy formation model. 
The contours represent high density regions (HDRs) of LAEs under
the same condition as \citet{H04}}
\label{LD}
\end{figure*}

\subsection{Basic Picture}
In Fig. \ref{gainen}, the schematic picture of the present
galaxy formation model is presented.
In the context of a conventional biased galaxy formation model, 
density peaks with the amplitude which 
exceeds a minimum threshold value
($\delta_{\rm{min}}$ in Fig. \ref{gainen}) in the linear regime
are identified as galaxies. 
In other words, only this threshold of fluctuations 
has been discussed as a parameter of biased galaxy formation \citep{Kau99, Ha04}.
%However it is difficult for this model to explain the observational 
%result of \citet{H04}. 
%Then, to explain this observational results of \citet{H04} without 
%contradicting $\Lambda$CDM theory, 

Here, we introduce an additional criterion with postulating
that LAEs evolve to galaxies with no strong
Ly$\alpha$ emission after their short lifetime.
More specifically, we take following assumption for LAEs:
(i) LAEs are galactic objects that form at peaks of
density fluctuations.
%the precursor of spheroidal systems such as elliptical galaxies 
%or galactic bulges in late-type galaxies.
%and such a system will form at early phase of galaxy evolution. 
(ii) LAEs are in the phase of their first starbursts. 
(iii) Chemical evolution of LAEs results in strong attenuation of Ly$\alpha$ 
emission due to the increase of dust,
and therefore cannot be observed as LAEs after their lifetime.
%In other word, chemical evolution of LAEs make them impossible 
%to be observed as LAEs, due to metal enrichment by themselves.
We incorporate this picture by setting 
a maximum threshold of density fluctuations
($\delta_{\rm{max}}$ in Fig. \ref{gainen}). 
Then, we regard the fluctuations between $\delta_{\rm{min}}$
and $\delta_{\rm{max}}$ as LAEs (a shaded region in Fig. \ref{gainen}). 
The growth time from $\delta_{\rm{min}}$ to $\delta_{\rm{max}}$
corresponds to the lifetime of LAEs. 
For instance, peak A in Fig. \ref{gainen} is the evolved galaxy 
that cannot be observed as LAEs because of exceeding $\delta_{\rm{max}}$ 
at the redshift. Peaks B and D can be observed 
as LAEs. Peak D is the oldest LAE. 
%Moreover, it is expected that LAEs such as peak D in Fig. \ref{gainen} 
%are discovered around large bias of galaxies such as peak A in Fig. \ref{gainen}.
%The two thresholds for the amplitude of density peaks, 
%$\delta_{\rm{min}}$ and $\delta_{\rm{max}}$, are 
%determined to match the number of observed
%LAEs and the two-point angular correlation function. 

\subsection{Numerical Method}
To compare our model with the observed clustering properties 
of LAEs \citep{H04},
we numerically generate LAE distributions,
and estimate the two-point angular correlation function
by following procedures. 

\subsubsection{Generation of LAE Spatial Distribution}
\label{GL}
It is assumed that the dynamical evolution of baryonic matter 
follows that of dark matter. 
Density fields of dark matter are created by generating random 
Gaussian density fields, and the dynamical evolution is represented 
by truncated Zel'dovich approximation \citep{Sa95}.
This approximation traces the growth of density fluctuations 
in the linear regime, and truncate nonlinear growth
by suppressing the amplitude of density fluctuations that becomes nonlinear.
In the present simulation, we use $k$-space Gaussian window 
$\Pi = \exp(- k^2 / 2k_{\rm{G}}^2)$ as truncation, 
where $k_{\rm G}$ corresponds to the scale that just enters nonlinear 
stage at a redshift $z$.
The truncated power spectrum of density fluctuation
at $z$, $P^*(k,z)$, is written as
\begin{equation}
P^*(k,z) = P(k,z)\Pi^2(k,z),
\end{equation}
where $P(k,z)$ is the power spectrum of density fluctuations
at $z$. The wavenumber $k_{\rm i}$ and real scale $r_{\rm i}$ have 
the relation of $k_{\rm i} = 2\pi / r_{\rm i}$. 

According to the $\Lambda$CDM theory, 
the physical size of 1$\sigma$ density fluctuation 
that collapses just at $z = 3.1$ is about $R = 1 h^{-1} \rm{~Mpc}$.
In this study, we consider density fluctuations down to this physical size. 
%This corresponds to considering a small density fluctuation than 
%standard galaxy formation model. 
%Therefore, we have to perform our numerical calculation with 
%a spatial resolution below this value.
In order to directly compare our model with the LAE data in
the comoving volume of $(50 h^{-1} \rm{~Mpc})^3$ \citep{H04}, 
we simulate the same comoving volume with $200^3$ grids.
The whole simulation box contains $4.5 \times 10^{15} M_{\odot}$ in 
dark matter component,
and each cell is $(0.25 h^{-1} \rm{~Mpc})^3$ 
and has $5.7 \times 10^{8} M_{\odot}$ on average. 
Next, we make coarse-graining of density fields 
by comoving volume of $(1 h^{-1} \rm{~Mpc})^3$ 
which corresponds to the physical size of interest. 
Each coarse-grained cell has $3.6 \times 10^{10} M_{\odot}$ 
in dark matter on average. The coarse-grained cells that satisfy the density 
fluctuation criterion, $\delta_{\rm{min}} \leq \delta \leq \delta_{\rm{max}}$,
are regarded as LAEs. 
The positions are assumed to be the center of mass in coarse-grained cell.

We choose several combinations of $\delta_{\rm{min}}$ and $\delta_{\rm{max}}$.
A set of $\delta_{\rm{min}}$ and $\delta_{\rm{max}}$ is constrained so that
the number of simulated LAEs should match
the observed number of LAEs at $z=3.1$ \citep{H04}.
Thus, if $\delta_{\rm{min}}$ is set, then $\delta_{\rm{max}}$ is determined
from the constraint of the number of LAEs.
In the linear regime, only density fluctuations with $\delta \geq 1.7$ corresponds
to collapsed objects \citep{Pea98}. Therefore, we consider $\delta_{\rm{min}}$
larger than 1.7. 
Resultant three-dimensional distributions of LAEs are projected into 
a two-dimensional plane to compare observed angular distributions. 

\subsubsection{Angular Correlation Function (ACF)}
\label{secACF}
To calculate the two-point ACF of the simulated spatial distribution of LAEs, 
we use a following well-known estimator,
\begin{equation}
w(\theta) = \frac{N_{\rm r}}{N_{\rm g}}\frac{\langle DD(\theta) \rangle}{\langle DR(\theta) \rangle} - 1,
\end{equation}
\citep{Peeb80, Pea98}, where $N_{\rm g}$ and $N_{\rm r}$ are the mean 
surface number density of simulated LAEs and 
that of randomly distributed points (RDPs), respectively. 
RDPs are distributed over the same area as LAEs.
$\langle DD(\theta) \rangle $ is the averaged pair number of LAEs in a range 
of $(\theta, \theta + d\theta)$, and $\langle DR(\theta) \rangle $ is
the averaged pair number between LAEs and RDPs in a range 
of $(\theta, \theta + d\theta)$. 
To raise the precision of statistics, we calculate ACFs
for 30 different realizations of density fluctuations and average them.
Then, the error on $w(\theta)$ is defined by 
the standard deviation of ACFs. 

\section{Results}
\subsection{Clustering Properties of LAEs}
In Fig. \ref{LD}, the spatial distributions of simulated LAEs 
are shown for different values of $\delta_{\rm{min}}$.
The contours depict `High Density Region' (HDR) defined 
under the same condition as \citet{H04},
where the number density smoothed with a Gaussian kernel of
$\sigma_{\rm G} = 90 \rm{~arcsec}$ 
(corresponding to $3{h^{-1}}\rm{~Mpc}$)
is equal to the mean number density in the entire field. 
The left panel in Fig. \ref{LD} is
the model with $\delta_{\rm{min}} = 1.7$ and $\delta_{\rm{max}} = 1.75$, 
and the middle panel is the model with
$\delta_{\rm{min}} = 2.5$ and $\delta_{\rm{max}} = 2.63$.
In the right panel, a conventional biased galaxy formation model 
is shown, where $\delta_{\rm{min}} = 5.4$ is assumed and 
all fluctuations with $\delta \geq \delta_{\rm{min}}$ are
regarded as LAEs.
In all these panels, we can recognize large-scale structures, but
the clustering manners are somewhat different. 
The spatial distributions in the biased galaxy formation model exhibit
very strong contrast and their clustered regions are fairly isolated. 
On the other hand, the distributions for 
$(\delta_{\rm{min}},\, \delta_{\rm{max}}) = (1.7,\, 1.75)$
or $(2.5,\, 2.63)$ appear to be belt-like and less clustered, 
similar to the observed spatial distribution of LAEs \citep{H04}. 

\begin{figure}
\includegraphics[width = 85mm]{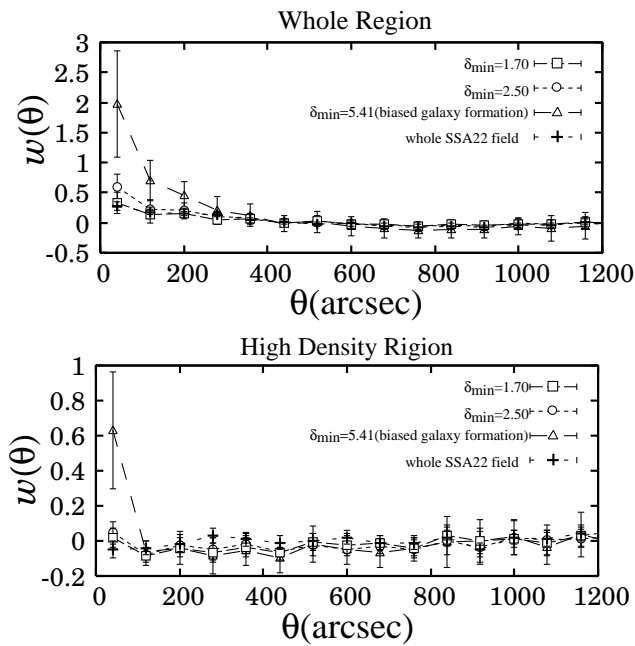}
\caption{Two-point angular correlation function (ACF) 
of LAE distributions for each model. 
Upper and lower panels show ACFs in the whole region and 
in HDR, respectively. 
Open squares represent ACFs for 
$\delta_{\rm{min}} = 1.7$ and $\delta_{\rm{max}} = 1.75$, 
and open circles for $\delta_{\rm{min}} = 2.5$ and 
$\delta_{\rm{max}} = 2.63$.
Open triangles represent ACFs for $\delta_{\rm{min}} = 5.4$ 
and $\delta_{\rm{max}} = \infty$, which is 
corresponding to a biased galaxy formation model. 
Crosses are ACFs of LAEs observed in SSA22a \citep{H04}. 
Note that the upper panel is different from the lower panel 
in the scale of vertical axis.}
\label{ACF}
\end{figure}

In order to quantify the difference in spatial distributions,
we calculate ACFs.
In Fig. \ref{ACF}, the resultant ACFs for all the model are presented.
Also, the ACF for LAEs observed in SSA22a field \citep{H04} is shown. 
The upper and the lower panels show the ACF in the whole region and 
that in HDR, respectively. The results show different behaviors
on scales smaller than $\sim 300 \rm{~arcsec}$. 
The biased galaxy formation model shows strong correlation on
small scales as expected in a standard biased model \citep{Kau99},
and obviously does not match the ACF of observed LAEs in SSA22a field. 
Furthermore, the model with 
$(\delta_{\rm{min}},\, \delta_{\rm{max}}) = (2.5,\, 2.63)$
results in slightly stronger ACF than the observation.
The model with $(\delta_{\rm{min}},\, \delta_{\rm{max}}) = (1.7,\, 1.75)$
remarkably agrees with the ACF of LAEs in SSA22a field.
In the HDR, the ACF exhibits negative correlation in the same way
as the observation. 
The reduction of ACF for smaller $\delta_{\rm{min}}$ is understood
as follows. In the random Gaussian density fields in
a $\Lambda$CDM universe, higher density peaks are more
clustered, while lower density peaks are located in 
less dense regions surrounding highest density regions. 
Thus, if small $\delta_{\rm{min}}$ is adopted and 
highest peaks are cut by $\delta_{\rm{max}}$, the objects of interest
are located in less dense regions and
accordingly the amplitude of ACF becomes smaller.
Hence, the result that the model with
$(\delta_{\rm{min}},\, \delta_{\rm{max}}) = (1.7,\, 1.75)$
reproduces the observed ACF implies that
LAEs at $z=3.1$ do not reside in highest density peaks, 
but are located in less dense regions.

Observationally, 
LAEs have been discovered around known overdensities that generally indicate 
strong correlation such as proto-cluster region including massive galaxies 
such as radio galaxies \citep{Ste00, H04, Ven05}. 
The observed overdensities may correspond to the situation shown
in the right pane of Fig. \ref{LD}, 
and the observed LAEs correspond to the left panel of Fig. \ref{LD}. 
In that sense, the results here look consistent with these observational 
features.
Hence, the picture in this paper can explain not only a correlation 
function but also other clustering properties of LAEs 
such as the morphology of HDR and the environment where LAEs 
at $z \sim 3$ is discovered. 
According to the recent large survey such as SDSS, 
ACFs of late-type galaxies show weaker correlation compared with that 
of early-type galaxies at $z \la 0.1$ \citep{Ze02}. 
That is to say, as well known, late-type galaxies are located 
at lower density fields. 
As shown here, LAEs at $z=3.1$ should be located in less dense regions.
Hence, it is suggested that a large fraction of LAEs at $z=3.1$
may be the precursors of late-type galaxies.

\subsection{Lifetime of LAEs}

As shown above, $\delta_{\rm{min}} = 1.7$ gives the best fit model
to account the observed ACF. Since $\delta = 1.7$ is a critical amplitude
for a fluctuation to collapse \citep{Pea98}, 
we can conclude that LAEs begin to shine just after the collapse. 
In other words, LAEs should be in the first phase of 
galaxy evolution. Since the model with 
$(\delta_{\rm{min}},\, \delta_{\rm{max}}) = (1.7,\, 1.75)$
agrees with the observed ACF at $z = 3.1$, 
LAEs are thought to shine during the growth time 
from $\delta_{\rm{min}}$ to $\delta_{\rm{max}}$. 
The fluctuation with $\delta_{\rm{max}}$ at $z = 3.1$ collapses
at a higher redshift $z_{\rm coll}$ when the amplitude
exceeds $\delta_{\rm{min}}$. 
Hence, the lifetime of LAEs can be assessed by the cosmic time
between $z = 3.1$ and $z_{\rm coll}$, which is $6.7 \times 10^7$~yr.
Here, there is a small uncertainty in this estimation. 
When $\delta_{\rm{min}}$ is chosen, $\delta_{\rm{max}}$ 
is determined to match the number of observed LAEs. 
Since we generate random numbers to produce density fluctuations, 
a different set of random numbers results in slight difference 
in $\delta_{\rm{max}}$. For the model of $\delta_{\rm{min}}=1.7$, 
we have $\delta_{\rm{max}}=1.75 \pm 0.01$ as a result of 30 different
realizations. Then, the lifetime of LAEs is estimated to be 
$(6.7 \pm 0.6) \times 10^7$~yr. Similarly, for the model of $\delta_{\rm{min}}=2.5$, 
we have $\delta_{\rm{max}}=2.63 \pm 0.02$. Then, 
the lifetime is slightly longer as $(2.0 \pm 0.4) \times 10^8$~yr. 

This result on LAE lifetime nicely agrees with an upper limit 
that is argued by realistic numerical simulations
for galactic evolution \citep{MU06a, MU06b}.
%Recently, \citet{MU06a, MU06b} have shown by 
%a high-resolution chemodynamical simulation of forming galaxies 
%that the spectral energy distribution (SED) exhibits conspicuous 
%Ly$\alpha$ emission at the earliest evolutionary phase of
%$< 3.0 \times 10^8$~yr. 
%Therefore, our estimate for the LAE lifetime is 
%consistent with the conclusion by \citet{MU06a, MU06b}.

\subsection{Luminosities of LAEs}

We also calculate Ly$\alpha$ luminosities of simulated LAEs, 
using an evolutionary spectral synthesis code 'PEGASE' \citep{PEGASE}. 
As a result, we have found that evaluated Ly$\alpha$ luminosities match
those of observed LAEs ($L_{\rm Ly\alpha} \sim 10^{42-43} \rm{erg s^{-1}}$) 
\citep{H04,Matsuda04,Breukelen05}. In this paper, 
density fields are coarse-grained by a scale of $1 h^{-1} \rm{~Mpc}$
which corresponds to 1$\sigma$ density fluctuations in 
the $\Lambda$CDM cosmology. If a smaller scale is taken, 
intrinsic Ly$\alpha$ luminosities fall short of $10^{42} \rm{erg s^{-1}}$
during Ly$\alpha$ bright phase. 
For instance, if a coarse-graining scale is $0.25h^{-1}$Mpc, 
intrinsic Ly$\alpha$ luminosities are $\sim 10^{41}~\rm{erg s^{-1}}$. 
On the other hand, if a scale larger than $1 h^{-1} \rm{~Mpc}$
is taken, the number of collapsed objects is not enough to
accout for the observed LAE number. 
Hence, 1$\sigma$ density fluctuations are favorable to explain 
the observations.

\subsection{ACF of LAEs at $3<z<6$}

\begin{figure}
\includegraphics[width = 85mm]{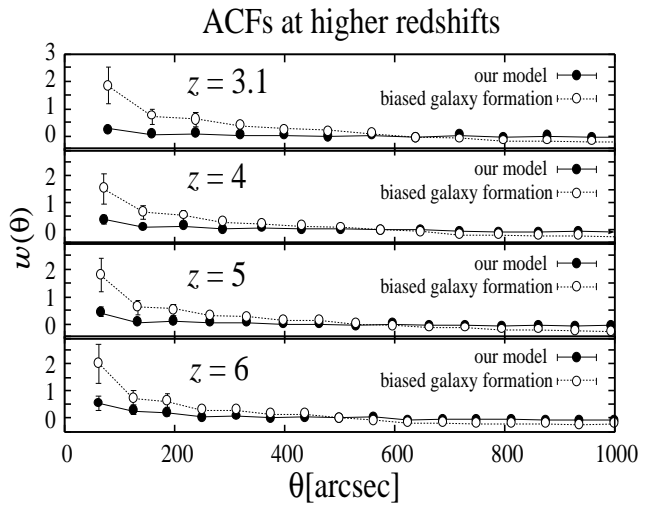}
\caption{Two-point angular correlation function (ACF) 
of simulated LAEs at redshifts at $z = 3.1$, 4.0, 5.0, and 6.0.
Open circles are short-lived LAE model, while 
filled circles are biased galaxy formation models. 
In the short-lived LAE model, density peaks between 
$\delta_{\rm{min}}=1.7$ and $\delta_{\rm{max}}=1.75$
are regarded as LAEs. 
In the biased galaxy formation model, the number of objects is 
scaled as to be the same as that in the short-lived LAE model.
So, $\delta_{\rm{min}}$ is set to $5.4$, $4.2$, $3.5$, and $2.8$,
at $z = 3.1$, $z = 4$, $z = 5$, and $z = 6$, respectively.} 
\label{zc}
\end{figure}

By assuming the best fit model 
($\delta_{\rm{min}} = 1.7$ and $\delta_{\rm{max}} = 1.75$),  
we can predict ACFs of LAEs at higher redshifts. 
In Fig. \ref{zc}, the prediction of ACFs at redshifts of 3.1, 4.0, 5.0, 
and 6.0 are presented. 
A biased galaxy formation model is
also presented, where the number of objects is 
scaled as to be the same as that in the best fit model.
%A value of $\delta_{\rm{min}}(\delta_{\rm{max}})$ of biased galaxy 
%formation model in each redshift are 
%$\delta_{\rm{min}} = 4.2(\mbox{or }\delta_{\rm{max}} = 13.9)$ at $z = 4$, 
%$\delta_{\rm{min}} = 3.5(\mbox{or }\delta_{\rm{max}} = 9.0)$ at $z = 5$, 
%and $\delta_{\rm{min}} = 2.8(\mbox{or }\delta_{\rm{max}} = 7.1)$ 
%at $z = 6$, respectively. 
$\delta_{\rm{min}}$ of biased galaxy formation model 
in each redshift is $\delta_{\rm{min}} = 4.2$ at $z = 4$, 
$\delta_{\rm{min}} = 3.5$ at $z = 5$, 
and $\delta_{\rm{min}} = 2.8$ at $z = 6$, respectively.
As seen in this figure, 
the ACF of best fit model approaches to that of biased galaxy formation model
at higher redshifts. 
In other words, a larger fraction of collapsed objects 
shine as LAEs at higher redshifts.
But, it is worth noting that
there is still noticeable difference between the best fit model and
a biased galaxy formation model even at $z = 6$.
It implies that a certain fraction has been already extinguished,
so that they are not detected as LAEs.

\section{Summary}

To account for the recently discovered large-scale structure of LAEs 
at $z=3.1$ \citep{H04},
we have introduced a novel picture for LAEs by focusing on
the lifetime of emitters.
We have simulated the spatial distributions of
collapsed objects by generating random Gaussian fluctuations
based on the truncated Zel'dovich approximation 
in the $\Lambda$CDM cosmology. 
We have found that a conventional biased galaxy formation model 
is not reconciled with the observed correlation function
of LAEs. If highest peaks above $\delta=1.75$ are cut and
mild peaks between $\delta=1.7$ and $\delta=1.75$ are
regarded as LAEs, the clustering properties including two-point
angular correlation function agree quite well with the observation.
Ly$\alpha$ luminosities also match those of observed LAEs.
The growth time from $\delta=1.7$ to $\delta=1.75$ can be
translated into the lifetime of LAEs, which is assessed to
be $(6.7 \pm 0.6) \times 10^7$~yr.
A fluctuation with $\delta=1.7$ corresponds to an object
that just collapses at the redshift. Thus, LAEs are thought to
be in the early evolutionary phase of galaxies, consistently
with a recent theoretical prediction \citep{MU06a, MU06b}.
We have also predicted the correlation function at redshift
higher than 3 in the picture of short-lived LAEs.
It is suggested that a certain fraction of young galaxies
have already ended the LAE phase even at redshift $z=6$. 

\section*{Acknowledgments}
We are grateful to T. Hayashino, Y. Matsuda and R. Yamauchi for offering 
valuable information and helpful comments. This work was supported in part 
by Grants-in-Aid, Specially Promoted Research 16002003 from MEXT in Japan.

%\appendix
%\section[]{Non}

\bsp

\label{lastpage}

\end{document}